# Cancer incidence estimation from mortality data: a validation study within a population-based cancer registry

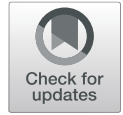

Daniel Redondo-Sánchez[1,2,3], Miguel Rodríguez-Barranco[1,2,3*] , Alberto Ameijide[4], Francisco Javier Alonso[5], Pablo Fernández-Navarro[3,6], Jose Juan Jiménez-Moleón[2,3,7] and María-José Sánchez[1,2,3,7]

## Abstract

**Background:** Population-based cancer registries are required to calculate cancer incidence in a geographical area, and several methods have been developed to obtain estimations of cancer incidence in areas not covered by a cancer registry. However, an extended analysis of those methods in order to confirm their validity is still needed.

**Methods:** We assessed the validity of one of the most frequently used methods to estimate cancer incidence, on the basis of cancer mortality data and the incidence-to-mortality ratio (IMR), the IMR method. Using the previous 15-year cancer mortality time series, we derived the expected yearly number of cancer cases in the period 2004–2013 for six cancer sites for each sex. Generalized linear mixed models, including a polynomial function for the year of death and smoothing splines for age, were adjusted. Models were fitted under a Bayesian framework based on Markov chain Monte Carlo methods. The IMR method was applied to five scenarios reflecting different assumptions regarding the behavior of the IMR. We compared incident cases estimated with the IMR method to observed cases diagnosed in 2004–2013 in Granada. A goodness-of-fit (GOF) indicator was formulated to determine the best estimation scenario.

**Results:** A total of 39,848 cancer incidence cases and 43,884 deaths due to cancer were included. The relative differences between the observed and predicted numbers of cancer cases were less than 10% for most cancer sites. The constant assumption for the IMR trend provided the best GOF for colon, rectal, lung, bladder, and stomach cancers in men and colon, rectum, breast, and corpus uteri in women. The linear assumption was better for lung and ovarian cancers in women and prostate cancer in men. In the best scenario, the mean absolute percentage error was 6% in men and 4% in women for overall cancer. Female breast cancer and prostate cancer obtained the worst GOF results in all scenarios.

**Conclusion:** A comparison with a historical time series of real data in a population-based cancer registry indicated that the IMR method is a valid tool for the estimation of cancer incidence. The goodness-of-fit indicator proposed can help select the best assumption for the IMR based on a statistical argument.

* Correspondence: miguel.rodriguez.barranco.easp@juntadeandalucia.es
[1]Granada Cancer Registry, Andalusian School of Public Health (EASP),
Campus Universitario de Cartuja, C/Cuesta del Observatorio 4, 18011
Granada, Spain
[2]Instituto de Investigación Biosanitaria de Granada (ibs.GRANADA), University
of Granada, Granada, Spain
Full list of author information is available at the end of the article

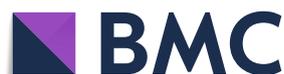

## Background

The cancer incidence rate is an essential epidemiological indicator in public health surveillance [1]. The calculation of cancer incidence requires a population-based cancer registry with all new cases of cancer in the region of reference, and given that population-based cancer registries follow international procedures, the reliability of their data in terms of exhaustiveness and validity is guaranteed [2, 3].

The reference publication of observed cancer incidence, *Cancer Incidence in Five Continents* (*CI5*), collects cancer incidence data from population-based cancer registries that comply with the quality standards established by the International Agency for Research on Cancer (IARC). Volume XI reports incidence data from 343 cancer registries in 65 countries for cancers diagnosed from 2008 to 2012 [4]. Despite the high number of data registries and participating countries, the population covered is less than 14.5% of the world population census of 2010 (the central point of the study period).

Although cancer is a worldwide public health problem and information regarding its incidence is necessary, there are regions not covered by a population-based cancer registry. In those regions, it is necessary to use other approaches to obtain incidence data that can be used in the healthcare planning and epidemiological surveillance of cancer. There are currently various initiatives that have provided estimations of cancer incidence for Europe and the other continents [5, 6]. In Spain, estimates have been published for 2015 for 25 cancer sites [7] within the framework of a research project of the Spanish Network of Cancer Registries (REDECAN) (https://redecan.org/redecan/es/index.html).

Some of the estimation methods used are very simple, whereas others are more complex and based on modeling techniques. The choice of method is based on the availability of the data (e.g., national incidence, national mortality, regional incidence) according to the health context of the region [8]. A comparison of the validity of several estimation methods was recently published [9]. However, this study was limited to only 1 year, and an extended analysis that goes beyond the estimation for a single value is still needed. This would make it possible to evaluate the stability of estimation methods, which could be affected by the time fluctuations of the data used.

The aim of this study was to evaluate the predictive performance of one of the most commonly used methods to derive cancer incidence rates from mortality and incidence-to-mortality ratio (IMR) data, by comparing expected cases with actual observed cases from the Granada Cancer Registry [10], based on a historical series. This method (IMR method) has been previously used to obtain incidence estimation for Spain over several time periods [7, 11, 12]. It was also the method used in the GLOBOCAN and EUCAN projects to estimate cancer incidence in Spain [8]. However, although the IMR method is being widely used, a validation based on a long time series has not yet been developed.

In addition, to further strengthen the method, we propose a measure of goodness-of-fit (GOF) to help select the best scenario for the predicted behavior of the IMR, which is a requirement for the application of this method.

## Methods

### Study population

This estimation method used data from Granada, a province in the southeast of Spain with a population of approximately 920,000 inhabitants in 2013 [13]. All deaths due to cancer between 1982 and 2010, according to sex, age group, and anatomical site, were obtained from the official death statistics of the Ministry of Health of the Government of Spain, which exhaustively collects all the medical certificates of death [14]. Population data between 1982 and 2017 was obtained from the Spanish National Institute of Statistics [13].

In addition, all incident cases diagnosed with cancer between 1985 and 2013 in the province of Granada were obtained from the population-based cancer registry of Granada. This registry complies with all quality standards for the publication of its data in the CI5 from volume VI (1985–1987) to volume XI (2008–2012), which is the most recent [4].

### Cancer sites

Our study focused on six frequent cancer sites for each sex as well as the total number of cancer incidence cases (except for non-melanoma skin cancer), based on the International Statistical Classification of Diseases and Related Health Problems: 10th revision [15]. More specifically, male cancers included stomach (C16), colon (C18), rectal (C19–C20), lung (C33–C34), prostate (C61), bladder (C67, D09.0, D41.4), and other sites except for non-melanoma skin (C00–C15, C17, C21–C32, C35–C43, C45–C60, C62–C66, C68–C96). Female cancers included colon (C18), rectal (C19–C20), lung (C33–C34), breast (C50), corpus uteri (C54), ovarian (C56), and other sites except for non-melanoma skin cancer (C00–C17, C21–C32, C35–C43, C45–49, C51–C53, C55, C57–C96). The total number of estimated cancer incidence cases (without non-melanoma skin cancer) was obtained by adding the number of estimated cases for all individual sites.

### Estimation method

The method applied in our study was an adapted version of the method using mortality data and mortality-to-incidence ratios [16]. The most recent estimates of cancer incidence in Spain were calculated with this method [7].

In the first stage of our study, we used the NORDPRED method [17] to estimate the number of deaths that occurred in the target year for each sex and anatomical site. This method uses age-period-cohort (APC) modeling based on a power link function to predict the number of deaths:

$$R_{ap} = \left(A_a + D \cdot p + P_p + C_c\right)^5 \tag{1}$$

where $R_{ap}$ is the mortality for a specific age range ($a$) during a specific period ($p$), $A_a$ is the age component of a specific age range ($a$), $D$ is the common drift parameter reflecting the linear component of the trend, $P_p$ is the nonlinear period component of a specific period ($p$), and $C_c$ is the nonlinear cohort component in a specific cohort ($c$).

In the second stage, the incidence-to-mortality ratio (IMR) of the target year was estimated for each sex and cancer type by means of a generalized linear mixed model (GLMM). The number of incident cases was the dependent variable, which was assumed to follow a Poisson distribution. The number of deaths was the offset of the model whereas the independent variables were age and year of cancer diagnosis. The effect of the year of diagnosis was analyzed by means of a second-degree polynomial function. The effect of age was smoothed by means of a linear spline with four nodes (10th percentile, first tertile, second tertile, and 90th percentile of the mortality pool).

The estimation of the parameters of the models was based on a Bayesian statistic approach using Markov chain Monte Carlo methods (MCMC) [7]. The projection until the target year was then obtained from these models, based on the assumed behavior of the IMR for each cancer/sex combination.

The statistical software R was used, with the functions of the NORDPRED method [17], and the BRugs package [18] with OpenBUGS software [19, 20].

### Assumptions regarding the mortality-to-incidence ratios

To obtain the estimation for the target year, it was necessary to make assumptions regarding the expected behavior of the IMR, based on the tendency observed and the knowledge of the cancer studied. This study calculated the estimated number of cases based on each of the following assumptions regarding IMR behavior, resulting in the following five scenarios:

C1: The IMR remains constant from the last available year.

C3: The mean of the last three available values of IMR remains constant until the target year.

C5: The mean of the last five available values of IMR remains constant until the target year.

L: The IMR has a linear trend that is projected onto the target year.

Q: IMR has a quadratic trend that is projected onto the target year.

Assumptions C3 and C5 are an extension of C1 in which greater robustness was sought for the constancy of the IMR, in order to minimize the effects of anecdotic fluctuations.

### Validity assessment

To assess the validity of the method, an iterative procedure was implemented to obtain the longest possible series of estimated values based on the mortality data. The objective was to compare them with real observed incidence data from the cancer registry.

The procedure aimed to reproduce the real-life context where the estimations would be obtained. Accordingly, in order to calculate the estimation for 2013 (the most recent year with available incidence data), we used a previous mortality series of 20 years (1991–2010) as well as a 3-year projection (2009–2013) and an IMR series of 15 years (1994–2008) with a 5-year projection (2009–2013) (see Additional file 3: Figure S1). This situation simulated the most realistic scenario, depending on the existing delay of the mortality records and the cancer records for the current year (2–3 years and 5 years, respectively, although this delay relies upon different factors, mainly legal and administrative contexts).

This procedure was repeated in a reverse year after year until 2004, based on the available mortality and incidence data in the study area (Additional file 3: Figure S1). The end result of this iterative procedure was a series of pairs of observed and estimated cases for each year of the period 2004–2013. For each pair of values, the relative deviation of the estimated cases was calculated with regard to the observed cases as:

$$100 \cdot \frac{\text{Expected} - \text{Observed}}{\text{Observed}} \tag{2}$$

Please note that when the relative deviation is negative, the number of observed cases was greater than the number of expected cases. Otherwise, the relative deviation is positive.

### Goodness-of-fit assessment

We used an indicator to assess GOF by comparing the estimated and observed cases. The aim was to summarize in a single value the global fit of the method across the time series by means of the mean absolute percentage error (MAPE) [21]. This GOF score was calculated and compared for each combination of sex and anatomical site and for all the scenarios for the assumed behavior of the IMR.

The indicator MAPE for any period starting in year $y_1$ and finishing in year $y_2$ is defined in the following way:

$$\text{MAPE}(s) = \frac{100}{(y_2 - y_1) + 1} \sum_{i=y_1}^{y_2} \left| \frac{E_i(s) - O_i}{O_i} \right| \qquad (3)$$

where $E_i(s)$ is the number of estimated cases for year $i$ under scenario $s$, and $O_i$ is the number of observed cases for year $i$.

This indicator reflects the deviation of the estimated values from the real observed values across the time series. Therefore, this indicator shows which assumption-based scenario is most suitable for each specific cancer site: scenario $s_1$ will be considered better than scenario $s_2$ when $\text{MAPE}(s_1) < \text{MAPE}(s_2)$.

## Results

This analysis included 43,884 cancer deaths that occurred between 1982 and 2010 in the province of Granada and 39,848 cancer incidence cases (23,197 in men and 16,651 in women) diagnosed between 2004 and 2013 in the province of Granada.

The linear assumption for the IMR trend provided the best GOF for prostate cancer in men, as well as for cancer of the lung, ovary, and other sites in women. The constant assumption (C1, C3, C5) was better for the rest of the sites in both sexes (Table 1 and Figs. 1 and 2). In all cases (except for male bladder cancer), projecting the mean of the last 3–5 years instead of the last IMR value improved the goodness-of-fit results. In contrast, the quadratic assumption had the worst GOF results in the cancer sites studied (Figs. 1 and 2). The GOF indicator for prostate cancer was worse in all scenarios (Fig. 1). In females, breast cancer and the "other" category also had worse results than the rest of the sites studied (Fig. 2).

In the best scenario for every combination of sex and anatomical site, the relative deviation of the number of estimated cases from the real cancer incidence was −0.3% in men and −0.5% in women (Table 1). For men, all the anatomical sites but the prostate were underestimated, and more balanced results were obtained for women, with overestimation for the corpus uteri, ovary, and others and underestimation in the colon, rectum, lung, and breast.

By specific cancer site, the relative deviation among men ranged from −9.1% for rectal cancer to 11.6% for prostate cancer during the entire period. For women, the relative deviation ranged from −12.6% for lung cancer to 14.3% for ovarian cancer. Considering the relationship between annual average cases and relative deviation, all anatomical sites except for the ovary, corpus uteri, and lung in women and rectum, other sites, and prostate cancer obtained results within the expected range, depending on the number of cases (Fig. 3).

Additional file 1: Table S1 and Additional file 2: Table S2 show the yearly variation percentages for all anatomical sites and scenarios. Additional file 4: Figure S2 and Additional file 5: Figure S3 show the observed and estimated

**Table 1** Observed and expected incidence cases in Granada 2004–2013 and goodness-of-fit measures for the best scenario

| Sex | Site | Observed cases | Expected cases | Relative deviation (%) | MAPE | Best scenario |
|---|---|---|---|---|---|---|
| Men | Colon | 1963 | 1888 | −3.8 | 8.63 | C3 |
| | Rectum | 1084 | 985 | −9.1 | 16.83 | C3 |
| | Lung | 3391 | 3339 | −1.5 | 4.46 | L |
| | Prostate | 4424 | 4936 | 11.6 | 27.37 | L |
| | Bladder | 2635 | 2594 | −1.6 | 8.48 | C1 |
| | Stomach | 801 | 760 | −5.1 | 14.09 | C5 |
| | Others | 8899 | 8624 | −3.1 | 3.53 | C3 |
| | All sites (except for non-melanoma skin) | 23,197 | 23,126 | −0.3 | 6.34 | |
| Women | Colon | 1467 | 1413 | −3.7 | 9.99 | C5 |
| | Rectum | 669 | 615 | −8.1 | 19.24 | C3 |
| | Lung | 557 | 487 | −12.6 | 18.26 | L |
| | Breast | 4220 | 4082 | −3.3 | 14.76 | C3 |
| | Corpus uteri | 1134 | 1262 | 11.3 | 17.35 | C5 |
| | Ovary | 615 | 703 | 14.3 | 18.89 | L |
| | Others | 7989 | 8012 | 0.3 | 8.68 | L |
| | All sites (except for non-melanoma skin) | 16,651 | 16,574 | −0.5 | 3.85 | |

*MAPE* mean absolute percentage error

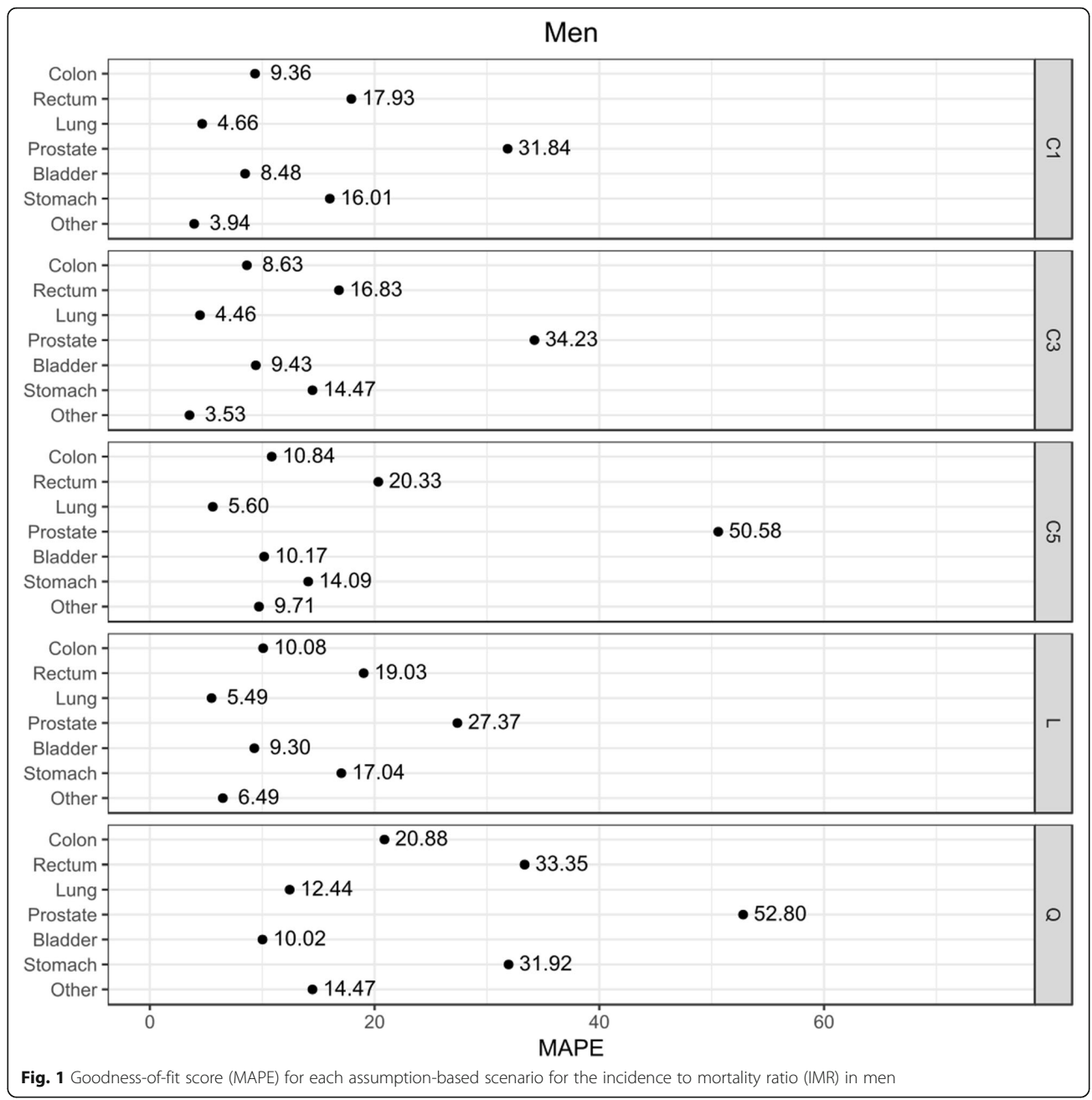

**Fig. 1** Goodness-of-fit score (MAPE) for each assumption-based scenario for the incidence to mortality ratio (IMR) in men

values for the best scenario. For men, lung cancer obtained the best approximations throughout the series with an annual relative deviation of 4.5% (ranging from −10.1 to 14.1%). In contrast, prostate cancer had an annual average deviation of 27.4%, which was as high as 136% in the last value of the series. For the rest of the sites, the fit was acceptable with annual mean deviations of a maximum of 16.8% (Additional file 1: Table S1 and Additional file 4: Figure S2). In the case of women, rectal cancer had the worst estimations, with deviations of 19.2%. Breast cancer shows an annual average deviation of 15%, although deviations rise above 20% in the last years of the series, with the expected cases underestimating the observed cases. The rest of the anatomical sites obtained average deviation percentages ranging from 10 to 20% (Additional file 2: Table S2 and Additional file 5: Figure S3).

Regarding the total number of cancer cases, the average annual deviation was −0.5% in both men and women. Except for the last year of the series, the relative deviation was lower than 8% in absolute terms. In 2013, this deviation increased to 25% for men and 5% for women, mainly due to the effect of prostate cancer and breast cancer, respectively (Table 2 and Fig. 4).

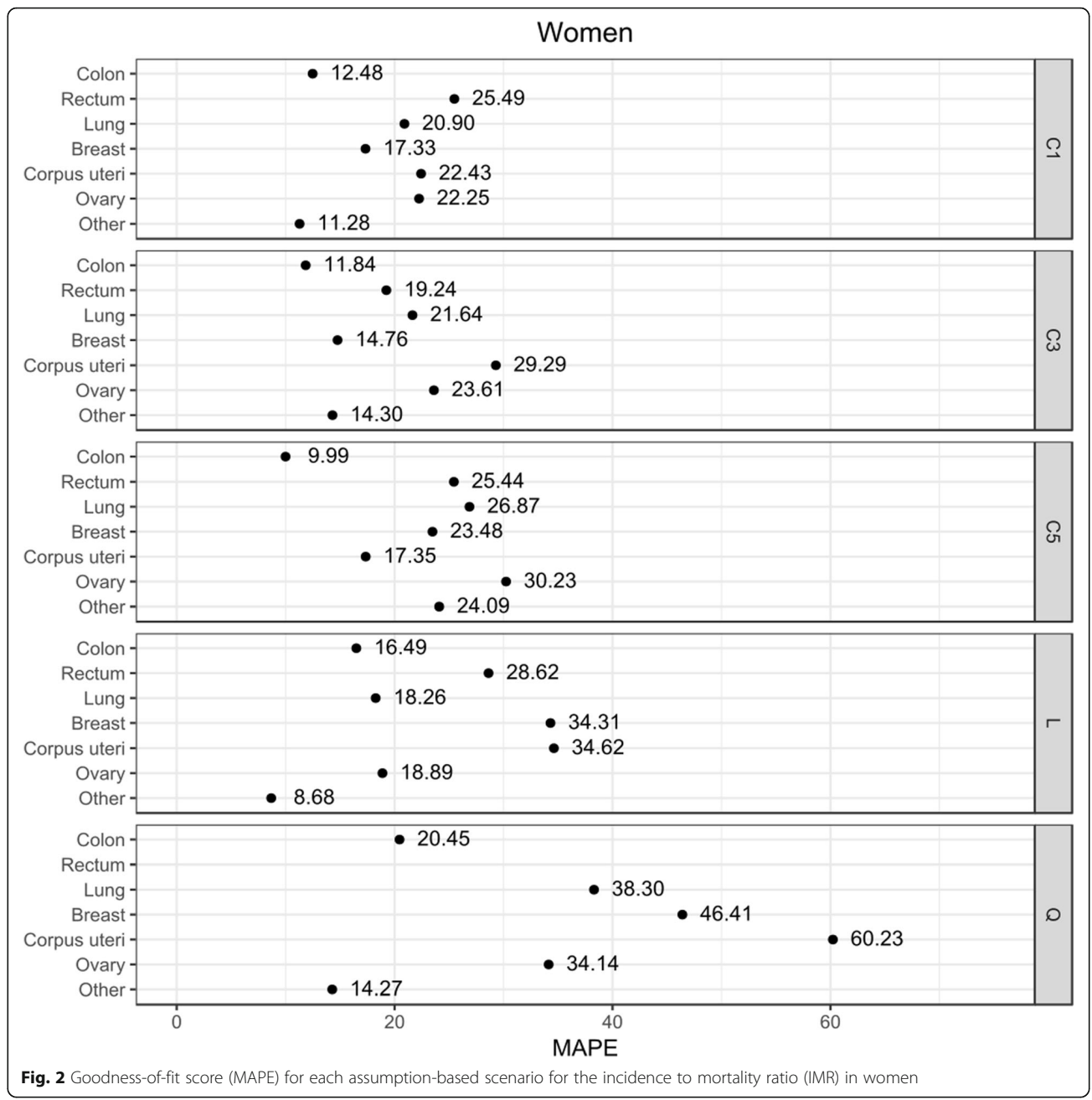

**Fig. 2** Goodness-of-fit score (MAPE) for each assumption-based scenario for the incidence to mortality ratio (IMR) in women

## Discussion

This research study analyzed one of the most frequently used methods for estimating cancer incidence data worldwide [6]. Although this method had been tested in previous research [9], it had never been validated with a long dataset over time. In addition, until now, the best hypothesis for the trend of the IMR had been selected solely on subjective criteria. To address this issue, we proposed an objective indicator to evaluate the goodness-of-fit in each scenario. This indicator will facilitate decision-making regarding the assumed behavior of the IMR when this method is applied in subsequent studies.

Our results showed that it is not advisable to use only 1 year for the IMR projection on the supposed constant because this value could be very high or very low in comparison with the normal trend and seriously bias the results. It was observed that a quadratic assumption did not work well for all sites in this analysis.

Similar to the national study, none of the scenarios examined seemed to fit IMR behavior with regard to breast cancer and especially prostate cancer. This was also reflected in the GOF indicator, which showed considerably higher scores for these sites. For these sites where no scenario was adequate, it is necessary to use new strategies to get suitable estimations and re-evaluate them using the same method. We believe that the IMR method has important limitations when it comes to cancers with sudden fluctuations of the IMR due to a public health intervention (e.g., a screening program, universal or opportunistic) [8, 9]. This should be taken into account for any other anatomical site that is affected by an early detection program in the study region.

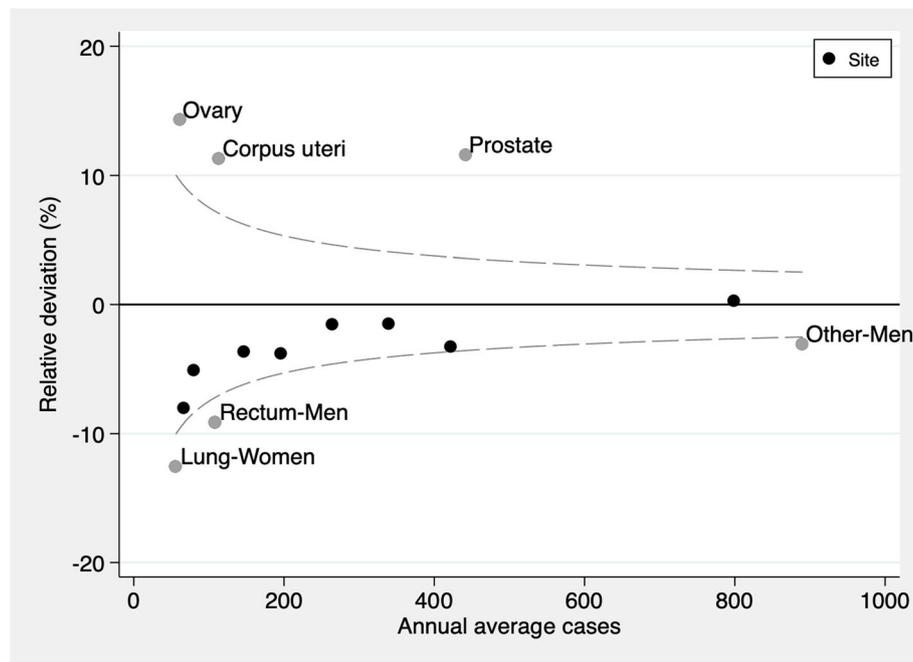

**Fig. 3** Funnel plot of the relative deviation of estimated incidence cases from the observed cases

In the best scenario, the results obtained in our study show that the method based on mortality data and IMR to derive cancer incidence cases provides good reliability for most cancer sites. When the time period was considered as a whole, the relative deviation between the estimated cases and real cases of cancer incidence in relation to the total cancer cases was −0.3% in men (23,126 expected cases vs. 23,197 observed cases) and −0.5% in women (16,574 expected cases vs. 16,651 observed cases). This low relative deviation was mainly caused because the relative deviations year by year were almost canceled out.

In a comparative study that contrasted 9 estimation methods with the number of actual cancer incidence cases diagnosed in 2010 in Norway, the deviation for the total number of cancer cases (using the same method) was 10.4% for men and 9.6% for women [9]. In our study, of the 14 sites studied, 7 showed absolute deviations lower than 5%, and 10 had absolute deviations lower than 10%. Only prostate cancer, corpus uteri cancer, ovarian cancer, and female lung cancer had values that exceeded this threshold. These results agree with those obtained by Antoni et al. [9], in which even higher deviations were obtained for these same sites.

**Table 2** Annual observed and expected incidence cases for all cancer sites (except non-melanoma skin) for the best scenario

| Year | Men | | | Women | | |
|---|---|---|---|---|---|---|
| | Observed cases | Expected cases | Relative deviation (%) | Observed cases | Expected cases | Relative deviation (%) |
| 2004 | 1990 | 1881 | −5.5 | 1459 | 1414 | −3.1 |
| 2005 | 2207 | 2100 | −4.8 | 1542 | 1510 | −2.1 |
| 2006 | 2144 | 2100 | −2.1 | 1510 | 1580 | 4.6 |
| 2007 | 2262 | 2123 | −6.1 | 1672 | 1575 | −5.8 |
| 2008 | 2363 | 2228 | −5.7 | 1635 | 1603 | −2.0 |
| 2009 | 2468 | 2277 | −7.7 | 1809 | 1725 | −4.6 |
| 2010 | 2370 | 2390 | 0.8 | 1739 | 1809 | 4.0 |
| 2011 | 2437 | 2381 | −2.3 | 1788 | 1716 | −4.0 |
| 2012 | 2527 | 2617 | 3.6 | 1695 | 1752 | 3.4 |
| 2013 | 2429 | 3029 | 24.7 | 1802 | 1890 | 4.9 |

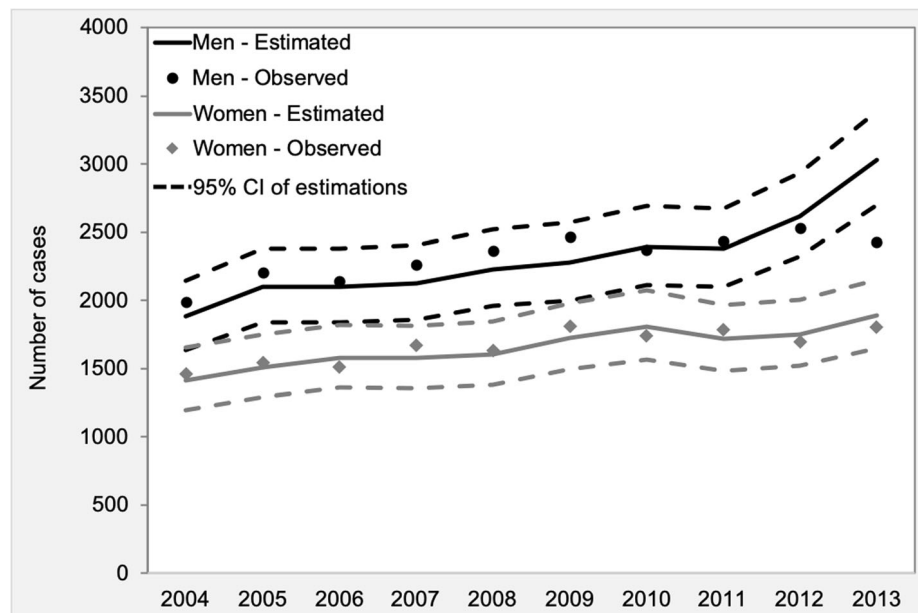

**Fig. 4** Observed incident cases vs. estimated cases using the best scenario for each site

Despite the limitations of this method in breast cancer (because of the screening program in the period of study, which affected IMR behavior), the relative deviation of the expected from the observed cases was only −1.1% overall (4175 expected cases and 4220 observed), even though values higher than 20% were obtained in the final years of the series. Prostate cancer showed a less satisfactory result since the average overestimation was 12%, and in the final year even reached 136%. These results may be due to overdiagnosis produced by opportunistic screening with PSA tests and confirm that the estimations for cancers affected by public health interventions should be interpreted with caution.

The method evaluated in this study has certain limitations. On the one hand, all methods that use mortality, survival, or IMR models are less precise when the number of cases or deaths is small, when the IMR is high, and/or when there are sudden changes in the IMR in relation to the period or age at diagnosis. For this reason, cancers of the corpus uteri, ovaries, and lungs in women obtained less satisfactory results than the other cancer types. Moreover, screening may produce sudden changes in the IMR in relation to age, time of diagnosis, and province [22].

However, the method analyzed in this research also had strengths. The use of recent data (3–5 years) allowed us to obtain accurate estimations by applying shorter projections, which minimized the risk of bias. In addition, the use of data based on official mortality statistics and included in population-based cancer registries guaranteed the quality of the information upon which the estimations were based. The possibility of applying different scenarios in the method made it sufficiently versatile so that it could be adapted to the specific characteristics of each cancer.

## Conclusions

The comparison for the first time over a historical time series between the obtained estimates and the real data observed in a population-based cancer registry demonstrated that the IMR method is a valid tool for cancer incidence estimation, as long as the data (incidence, mortality, and population) is of high quality. The optimal scenarios proposed in this work could be used for the application of the IMR method in other regions with similar characteristics, such as the health system, the expected survival of patients, or the risk factors of cancer in the reference population. The valuable work of population-based cancer registries, which collect exhaustive incidence data, together with the application of statistical techniques that are used to obtain both recent and accurate estimations, will provide users with crucial data for the surveillance of cancer in any region of the world.

## Supplementary Information

---

**Additional file 1: Table S1.** Observed cases and expected cases under each scenario, and goodness of fit measures. Men.

**Additional file 2: Table S2.** Observed cases and expected cases under each scenario, and goodness of fit measures. Women.

**Additional file 3: Fig. S1.** Design of the data used in the iterative procedure to derived the incidence time series.

**Additional file 4: Fig. S2.** Number of observed and expected (with 95% CI) cases under the best scenario for each site. Men.

**Additional file 5: Fig. S3.** Number of observed and expected (with 95% CI) cases under the best scenario for each site. Women.

---

### Acknowledgements
We would like to thank the staff of the Granada Cancer Registry for their work in the collection and registration of cancer cases.
The views expressed by the authors from Carlos III Institute of Health are those of them and not necessarily those of the Carlos III Institute of Health.

### Authors' contributions
M. Rodríguez-Barranco conceived the study and drafted the original manuscript. D. Redondo-Sanchez prepared the data and developed the analysis. A. Ameijide formulated the code to run the method. F. J. Alonso contributed to the assessment of the goodness-of-fit analysis. M. J. Sánchez provided anonymized access to the cases from the Granada Cancer Registry. All authors helped to write the text and approved the final manuscript.

### Funding
This research was supported with the subprogram "Cancer surveillance" of the CIBER of Epidemiology and Public Health (CIBERESP). This work has been also partially supported by grant PGC2018-098860-B-I00 (MINECO/FEDER). M. J. Sánchez is supported by the Andalusian Department of Health Research, Development and Innovation, project grant PI-0152/2017.

### Availability of data and materials
The datasets of aggregated mortality and IMR are available from the corresponding author upon reasonable request.

### Declarations

#### Ethics approval and consent to participate
Not applicable.

#### Consent for publication
Not applicable.

#### Competing interests
The authors declare that they have no competing interests.

### Author details
¹Granada Cancer Registry, Andalusian School of Public Health (EASP), Campus Universitario de Cartuja, C/Cuesta del Observatorio 4, 18011 Granada, Spain. ²Instituto de Investigación Biosanitaria de Granada (ibs.GRANADA), University of Granada, Granada, Spain. ³CIBER of Epidemiology and Public Health (CIBERESP), Madrid, Spain. ⁴Tarragona Cancer Registry, Foundation Society for Cancer Research and Prevention (FUNCA), Pere Virgili Health Research Institute (IISPV), Reus, Spain. ⁵Department of Statistics, Faculty of Sciences, University of Granada, Granada, Spain. ⁶Cancer and Environmental Epidemiology Unit, National Center for Epidemiology, Carlos III Institute of Health, Madrid, Spain. ⁷Department of Preventive Medicine and Public Health, University of Granada, Granada, Spain.

**Publisher's Note**

Springer Nature remains neutral with regard to jurisdictional claims in published maps and institutional affiliations.